\documentclass[aps,reprint,prl,superscriptaddress,showpacs,floats,floatfix]{revtex4-1}
\usepackage{graphicx}
\usepackage{amsmath,graphicx,dcolumn}
\usepackage{hyperref}
\usepackage[usenames]{color}
\usepackage{datetime}
\usepackage{textcomp}
\usepackage{ulem}
\usepackage{morefloats}
\usepackage{mathptmx}
\usepackage{indentfirst}
\usepackage{bm}
\usepackage[cmyk]{xcolor}
\usepackage{siunitx}

\newcommand{\blue}{\textcolor{blue}}

\newcommand*{\EIA}{EuIn$_2$As$_2$}
\newcommand*{\mjcm}{$\mu$J/cm$^2$}

\begin{document}

\title{Exploring Intrinsic Magnetic Topological Insulators: The Case of EuIn$_2$As$_2$}

\author{Hao Liu}
\affiliation{School of Physics, Central South University, Changsha 410083, Hunan, China}

\author{Qi-Yi Wu}
\affiliation{School of Physics, Central South University, Changsha 410083, Hunan, China}

\author{Chen Zhang}
\affiliation{School of Physics, Central South University, Changsha 410083, Hunan, China}

\author{Jie Pang}
\affiliation{Beijing National Laboratory for Condensed Matter Physics, Institute of Physics, Chinese Academy of Sciences, Beijing 100190, China}
\affiliation{School of Physical Sciences, University of Chinese Academy of Sciences, Beijing 100049, China}

\author{Bo Chen}
\affiliation{School of Physics, Central South University, Changsha 410083, Hunan, China}

\author{Jiao-Jiao Song}
\affiliation{School of Physics, Central South University, Changsha 410083, Hunan, China}

\author{Yu-Xia Duan}
\affiliation{School of Physics, Central South University, Changsha 410083, Hunan, China}

\author{Ya-Hua Yuan}
\affiliation{School of Physics, Central South University, Changsha 410083, Hunan, China}

\author{Hai-Yun Liu}
\affiliation{Beijing Academy of Quantum Information Sciences, Beijing 100085, China}

\author{Chuan-Cun Shu}
\affiliation{School of Physics, Central South University, Changsha 410083, Hunan, China}

\author{Yuan-Feng Xu}
\affiliation{Center for Correlated Matter and School of Physics, Zhejiang University, Hangzhou 310058, China}

\author{You-Guo Shi}
\affiliation{Beijing National Laboratory for Condensed Matter Physics, Institute of Physics, Chinese Academy of Sciences, Beijing 100190, China}
\affiliation{School of Physical Sciences, University of Chinese Academy of Sciences, Beijing 100049, China}

\author{Jian-Qiao Meng}
\email{Corresponding author: jqmeng@csu.edu.cn}\affiliation{School of Physics, Central South University, Changsha 410083, Hunan, China}

\date{\today}

\begin{abstract}

In this paper, ultrafast optical spectroscopy was employed to elucidate the intricate topological features of EuIn$_2$As$_2$, a promising candidate for a magnetic topological crystalline axion insulator. Our investigation has revealed the exceptional sensitivity of the ultrafast carrier dynamics to the antiferromagnetic phase transition ($T_N$ $\approx$ 16 K). Below $T_N$, two distinct, extremely low-energy collective modes, $\omega_1$ and $\omega_2$, with frequencies of $\sim$9.9 and 21.6 GHz at $T$ = 4 K, respectively, were observed, exhibiting strong temperature dependence. The higher-frequency mode is a magnon, and the lower-frequency mode, whose origin is currently uncertain, cannot be attributed to a coherent acoustic phonon. Additionally, high-fluence photoexcitation suggests the possibility of light-induced nonthermal phase transitions, such as the disruption of antiferromagnetic order. These findings provide insights into the interplay between topology and magnetism in EuIn$_2$As$_2$, making it a promising candidate for applications in quantum information and spintronics.

\end{abstract}


\maketitle

\section{I. INTRODUCTION}
\vspace*{-0.3cm}
Topological insulators, a revolutionary class of quantum materials, manifest topological surface states within an insulating bulk \cite{MZHasan2010, SCZhang2011}. These surface states, protected by time-reversal symmetry (TRS), exhibit linear dispersion near Dirac points, rendering them impervious to non-magnetic perturbations. The introduction of additional magnetic order, disrupting TRS, results in magnetic topological materials with a unique band topology, promising distinct quantum phenomena \cite{Bernevig2022, YFXu2020}. Recent theoretical predictions highlight EuIn$_2$As$_2$ as a compelling candidate \cite{YFXu2019}, boasting intrinsic magnetic order and a three-dimensional hexagonal structure.

EuIn$_2$As$_2$ possesses a layered centrosymmetric crystal structure (space group $P6_3$/$mmc$, No. 194) with alternating stacking of Eu$^{2+}$ and [In$_2$As$_2$]$^{2-}$ layers along the $c$ axis, undergoing a paramagnetic (PM) to antiferromagnetic (AFM) transition at the N\'{e}el temperature $T_N$ $\approx$ 16 K \cite{Goforth2008, Rosa2012}. Theoretical frameworks propose {\EIA} as a potential topologically nontrivial magnetic insulator, capable of hosting AFM axion insulator phases \cite{YFXu2019, Riberolles2021, JRSoh2023, SRegmi2020}. Advances in theoretical calculations underscore the critical influence of magnetic moment orientation on emergent topological states, predicting a topological crystalline insulator phase with some gapless surface states emerging on (100), (010), and (001) surfaces for in-plane oriented magnetic moments \cite{YFXu2019, SRegmi2020, TSato2020} and a higher-order topological insulator phase with gapped surface states emerging on (100) and (001) surfaces for out-of-plane oriented magnetic moments \cite{YFXu2019}. Unlike previously discovered material systems,  density functional theory (DFT) calculations reveal that topological surface states in the stoichiometric form of {\EIA} are precisely located at the Fermi energy ($E_F$).

Despite progress in understanding, controversies persist, particularly regarding the nature of the axion-insulator phase, such as whether there is a surface state energy gap associated with TRS breaking. Previous magnetic measurements indicate $A$-type AFM order in {\EIA} \cite{Goforth2008, Rosa2012}, but recent neutron diffraction \cite{Riberolles2021} and resonant elastic x-ray scattering \cite{JRSoh2023} experiments suggest a low-symmetry helical AFM order, also satisfying the symmetry requirements for an axion insulator.

Magnetotransport measurements reveal negative magneto resistance, indicating robust spin scattering by localized magnetic moments \cite{Goforth2008, YZhang2020}. Scanning tunneling microscopy and spectroscopy (STM/STS) data show a partial surface state gap ($\sim$40 meV) below $T_N$, diminishing with increasing temperatures but remaining finite above $T_N$ \cite{MDGong2022}. Angle-resolved photoemission spectroscopy (ARPES) confirms hole-type Fermi pockets \cite{SRegmi2020, TSato2020, YZhang2020}, a heavily hole-doped surface state, and inverted bulk bands in the AFM state \cite{TSato2020}, suggesting topological surface states are located above $E_F$. Notably, ARPES confirms the dominance of surface states near $E_F$, but fails to detect magnetic exchange gaps within them \cite{SRegmi2020, TSato2020, YZhang2020}.

Ultrafast optical spectroscopy serves as a powerful tool for probing the intricate low-energy electron dynamics in correlated materials \cite{DNBasov2011}, offering insights into the complex behaviors of systems such as transition metal dichalcogenides \cite{SXZhu2021, JBQi2010}, high-temperature superconductors \cite{CZhang2022, QYWu2023A, QYWu2023B, Kabanov1999, Demsar2006}, and heavy fermions \cite{YZZhao2023, KSBurch2008, YPLiu2020, JQi2013}. Sensitivity to changes in the low-energy electronic structure, especially the opening of narrow energy gaps near the $E_F$ that may lead to bottleneck effects \cite{CZhang2022, QYWu2023A, QYWu2023B, Kabanov1999, Demsar2006, YZZhao2023, KSBurch2008, YPLiu2020, JQi2013, Rothwarf1967}, renders it particularly valuable. Additionally, ultrafast optical spectroscopy excels in detecting collective bosonic excitations, providing a reliable indicator for phase transitions \cite{KSBurch2008, YPLiu2020, YZZhao2023}. While ultrafast optical spectroscopy probes a material's bulk to a certain depth, it can offer valuable insights into the dynamics of surface states under specific conditions,  particularly when surface states dominate the electronic structure near $E_F$, such as topological insulators \cite{VIyer2018, PSharma2022}.

\begin{figure*}[tbp]
\begin{center}
\includegraphics[width=1.7\columnwidth]{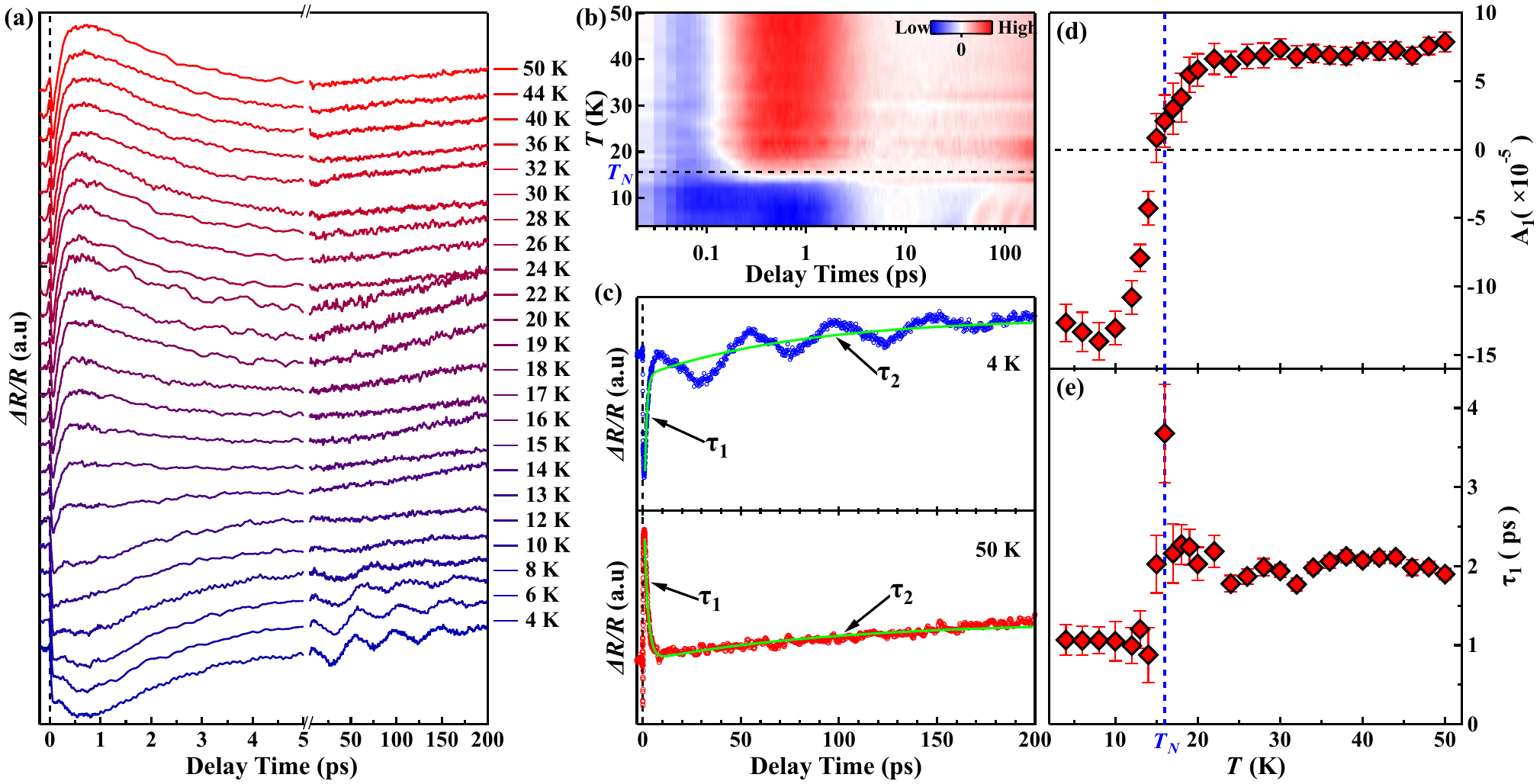}
\end{center}
\vspace*{-0.5cm}
\caption{(a) $\Delta R/R$ as a function of delay time over a temperature range from 4 to 50 K at a pump fluence of 53 {\mjcm}. Note the break in the $x$ axis. (b) Two-dimensional pseudocolor map of $\Delta R/R$ as a function of temperature and delay time. (c) Illustration of fitting using Eq. (\blue{1}) (upper panel: 4 K; bottom panel: 50 K). The arrows indicate the corresponding relaxation process. (d) and (e) $T$ dependence of $A_1$ and $\tau_1$, respectively.}\label{FIG:1}
\end{figure*}

Our study utilizes ultrafast optical spectroscopy to investigate the intricate topological states of {\EIA}, aiming to address current controversies and provide dynamic insights. Employing ultrafast techniques, we seek to unveil the real-time evolution of topological features, offering a nuanced understanding of the interplay between magnetism and topology in this fascinating material. The quasiparticle relaxation in {\EIA} undergoes a dramatic change in the vicinity of $T_N$. Below $T_N$, our measurements show the emergence of two extremely low-energy collective modes, $\omega_1$ and $\omega_2$, with frequencies of $\sim$9.9 and 21.6 GHz at $T$ = 4 K, respectively, both exhibiting a strong temperature dependence. The temperature- and fluence-dependent measurements suggest that the origin of $\omega_1$ remains unclear and cannot be attributed to a coherent acoustic phonon, while $\omega_2$ is associated with a magnon. Moreover, our high-fluence photoexcitation results suggest the potential for photoinduced nonthermal phase transitions. Our findings firmly establish {\EIA} as a promising material for investigating the interplay between topology and magnetism.
\vspace*{-0.3cm}

\section{II. EXPERIMENTAL DETAILS}
\vspace*{-0.3cm}

High-quality single crystals of {\EIA} were grown utilizing the self-flux method \cite{YLi2021}, demonstrating a $T_N$ of approximately 16 K (refer to Fig. \blue{S1} in the Supplemental Material for the magnetic measurements \cite{SupplementalMaterial}). We performed ultrafast time-resolved differential reflectivity measurements ($\varDelta R/R$) at a center wavelength of 800 nm ($\sim$1.55 eV) using a 1 MHz Yb-based femtosecond (fs) laser oscillator \cite{CZhang2022}. These measurements were carried out on a freshly cleaved (001) surface under a high vacuum of $10^{-6}$ mbar. The pump and probe beams were focused on the sample at nearly normal incidence, with spot diameters of approximately 85 $\mu$m for the pump beam and 40 $\mu$m for the probe beam. The pump and probe beams were $s$ and $p$ polarized, respectively.

\vspace*{-0.3cm}
\section{III. RESULTS AND DISCUSSIONS}
\vspace*{-0.3cm}

Figure \blue{1(a)} presents the differential reflectivity $\Delta R/R$ of {\EIA} at various temperatures (4-50 K) at a pump fluence of $\sim$53 {\mjcm}. Photoexcitation induces immediate changes in $\Delta R/R$, followed by multiple recovery processes with different lifetimes, indicating diverse relaxation pathways. The curves exhibit a pronounced temperature dependence. Notably, the peak around 0.7 ps transforms into a valley with decreasing temperature.
Below $T_N$, damped oscillations appear after a few picoseconds. These observations contrast with the minimal temperature dependence reported by Wu $et$ $al$. \cite{QiongWu2023}. The discrepancy likely stems from the different pump wavelengths (800 nm vs. 400 nm). Above $T_N$, optical conductivity suggests a transition from the valence to the conduction band at a wave number corresponding to 800 nm. Below $T_N$, the AFM transition splits the valence band \cite{BXu2021}. Therefore, transient reflectivity using an 800 nm pump wavelength offers a more sensitive probe of the AFM transition in {\EIA} compared to the approach employed by Wu $et$ $al$. \cite{QiongWu2023}.

Figure \blue{1(b)} presents a two-dimensional (2D) map of $\Delta R/R$ (pump-probe delay versus temperature). The initial relaxation ($\sim$0.3 ps) reveals pronounced variations in $\Delta R/R$ near $T_N$, with a clear positive-to-negative signal transition. This behavior confirms $\Delta R/R$ effectiveness in capturing electronic structure changes during the AFM transition. It aligns with expectations \cite{YZZhao2023} and supports prior findings from ARPES \cite{SRegmi2020, TSato2020} and infrared spectroscopy \cite{BXu2021} of significant alterations in the electronic structure during the AFM transition.

A quantitative analysis of quasiparticle dynamics was conducted to investigate its temperature-dependent behavior. We focus here on recombination processes occurring after electron-electron and electron-phonon thermalization, when carriers have already relaxed to the vicinity of the $E_F$ \cite{HLi2019}. The solid green lines in Fig. \blue{1(c)} suggests that the nonoscillatory response over a considerable time domain ($\sim$1-200 ps) fits well with a bi-exponential decay, using the expression
\begin{equation}
\begin{split}
\frac{R(t)}{R} = \sum\limits_{i=1,2} A_{i}{\rm exp}(-\frac{t-t_0}{\tau_{i}}) + C
\end{split}
\end{equation}
where $A_{i}$ is the amplitude and $\tau_{i}$ is the relaxation time of the \textit{i}th nonoscillatory signal which describes carrier dynamics. $C$ is a constant representing long-lifetime processes.

Figures \blue{1(d)} and \blue{1(e)} show the temperature dependence of extracted amplitude ($A_1$) and relaxation time ($\tau_1$). Both exhibit anomalies near $T_N$, reflecting the influence of AFM order on electron-hole recombination \cite{Majchrzak2023}. $A_1$ displays a slow decrease followed by a rapid drop and sign reversal near $T_N$ before saturating at lower temperatures. Similarly, the relaxation time $\tau_1$ exhibits a sharp change around $T_N$. These behaviors, reminiscent of those observed in superconductors \cite{Kabanov1999, Demsar2006, CZhang2022, QYWu2023A, QYWu2023B} and heavy fermion compounds \cite{JQi2013, KSBurch2008, YPLiu2020, YZZhao2023}, have been attributed to the opening of a narrow energy gap in the density of states (DOS) near $E_F$. STM/STS measurements \cite{MDGong2022} and DFT calculations \cite{SRegmi2020, Riberolles2021} show that the spin-orbit coupling-induced bulk gap varies little across $T_N$ and persists above it. Furthermore, theoretical studies suggest that the AFM transition may open a gap at the Dirac cone of the surface state, regardless of whether the AFM order is out of plane \cite{YFXu2019, SRegmi2020} or helical \cite{Riberolles2021}. Notably, the sign inversion of $A_1$ near $T_N$ introduces significant uncertainties in the fitting process for $A_1$ and $\tau_1$. Therefore, our data are insufficient to conclusively support the existence of this gap. Thus, we attribute the anomalies in both $A_1$ and $\tau_1$ to changes in the bulk electronic structure \cite{BXu2021}, rather than the opening of a narrow bulk or surface magnetic gap. Future work using time-resolved and angle-resolved photoemission spectroscopy (TR-ARPES) will be crucial to precisely characterize the unoccupied band structure, particularly its temperature dependence and the gap position relative to $E_F$.

\begin{figure}[tbp]
\begin{center}
\includegraphics[width=0.98\columnwidth]{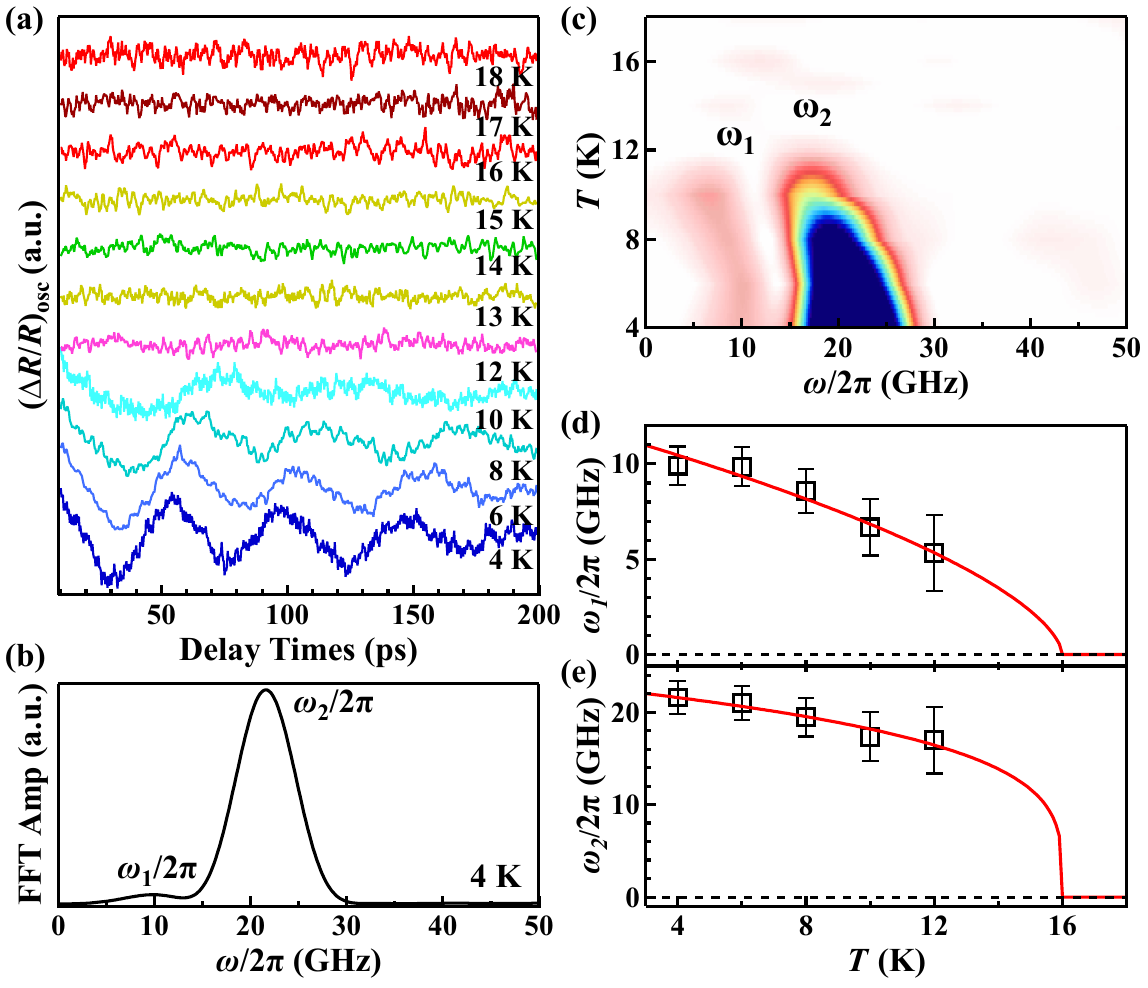}
\end{center}
\vspace*{-0.5cm}
\caption{(a) Extracted oscillations over a temperature range from 4 to 18 K at a pump fluence of 53 {\mjcm}. Curves are shifted for clarity. (b) Fast Fourier transform (FFT) frequency-domain data at 4 K. (c) FFT spectrum color intensity map as a function of frequency and temperature. (d) and (e) The derived frequencies of $\omega_1$ and $\omega_2$ as a function of temperature. The red lines are fits described in the main text.}\label{FIG:3}
\end{figure}

Following the low-fluence analysis, we investigated low-frequency collective bosonic excitations, known indicators of phase transitions \cite{KSBurch2008, YZZhao2023, YSWang2023}. Figure \blue{2(a)} depicts time-domain oscillations extracted from the low-temperature transient reflectivity data [Fig. \blue{1(a)}] after background removal. The oscillation periods and amplitudes vary with temperature, increasing and decreasing, respectively. Notably, the oscillations are exclusively observed below $T_N$ and vanish above it.

Oscillatory components were extracted by performing fast Fourier transform (FFT). Figure \blue{2(b)} presents two pronounced extremely low-energy modes, $\omega_1$ and $\omega_2$, at frequencies of $\sim$9.9 (i.e., 0.04 meV or 0.33 cm$^{-1}$) and $\sim$21.6 GHz  (i.e., 0.09 meV or 0.71 cm$^{-1}$) at a temperature of 4 K. Figure \blue{2(c)} displays the FFT spectrum as a function of frequency and temperature. The two gigahertz modes, $\omega_1$ and $\omega_2$, were exclusively observed at temperatures lower than $T_N$, with both frequencies consistently diminishing and experiencing softening as $T_N$ is approached. The presence of the two extremely low-energy modes is limited to temperatures below $T_N$, suggesting they are either of magnetic origin or highly responsive to the magnetic ordering.

The extracted temperature dependence of $\omega_1$ and $\omega_2$ frequencies is plotted in Figs. \blue{2(d)} and \blue{2(e)}, respectively. The temperature dependence of $\omega_1$ and $\omega_2$ can be effectively fitted using a mean-field-like $T$ dependence \cite{RVYusupov2008}, $\omega$ = $\omega$(0)$(1-T/T_N)^{2\beta}$, where $T_N$ = 16 K and $\beta$ are the N\'{e}el temperature and critical exponent, respectively. This fitting yields zero-temperature frequencies of $\omega_1$(0) $\simeq$ 12.5 GHz and $\omega_2$(0) $\simeq$ 23.2 GHz, with corresponding $\beta$ values of $\beta_{\omega_1}$ $\simeq$ 0.31 and $\beta_{\omega_2}$ $\simeq$ 0.123, respectively. 2$\beta_{\omega_2}$ matches the expected value of 0.25 for a two-dimensional Ising system \cite{MFCollins1989}. These results are consistent with the magnetic susceptibility measurements, suggesting that {\EIA} is a two-dimensional antiferromagnet \cite{Rosa2012, YZhang2020}. This noteworthy observation unequivocally confirms the intrinsically 2D character of {\EIA}'s magnetic ordering, which has been observed in other antiferromagnets, such as the two-dimensional antiferromagnet NiPS$_3$ \cite{Afanasiev2021}. At a pump fluence of 180 {\mjcm}, Wu $et$ $al$.'s time-resolved magneto-optical Kerr effect (TR-MOKE) measurements have revealed an 18 GHz spin precession mode under low magnetic fields, disappearing above $T_N$, while a persistent 35 GHz coherent acoustic phonon mode persists up to 25 K in high magnetic fields \cite{QiongWu2023}.

Considering that our measurements were conducted without the influence of a magnetic field, and given that the $\omega_2$ mode disappears after a magnetic phase transition induced by intense pump fluence (to be discussed in detail later), we infer that the $\omega_2$ mode originates from magnetic order, signifying its nature as a magnon. The localized Eu 4$f$ spin, positioned approximately 1.7 eV below $E_F$ and energetically separated from the initial optical excitation (1.55 eV), is effectively probed by ultrafast optical spectroscopy due to its strong interatomic exchange coupling with bulk topological bands. In EuIn$_2$As$_2$, below $T_N$, electrons excited from In 5$s$ and As 4$p$ orbitals decay to the ground state via two primary mechanisms: electron-phonon coupling-induced spin flips and spin-orbit coupling-induced spin flips. Consequently, the localized Eu 4$f$ spins undergo disorder through coupling with spin-flipped $sp$ orbital electrons, leading to coherent magnon emission to conserve angular momentum. This process, akin to impulsive stimulated Raman scattering (ISRS) via spin-flip scattering, resembles an ultrafast inverse Cotton-Mouton effect, where the linearly polarized excitation pulse acts on spins as a short effective field pulse \cite{Kalashnikova2007, Afanasiev2021, Nishitani2013}. Ultrafast optical spectroscopy functions as a powerful tool for characterizing nonequilibrium spin dynamics and their intricate interplay with electrons and the lattice. The temperature evolution of the lower-frequency oscillation $\omega_1$ differs, exhibiting distinct critical parameters compared to $\omega_2$. However, with the above dataset, we are unable to ascertain the origin of $\omega_1$.

\begin{figure}[tbp]
\centering
\includegraphics[width=0.98\columnwidth]{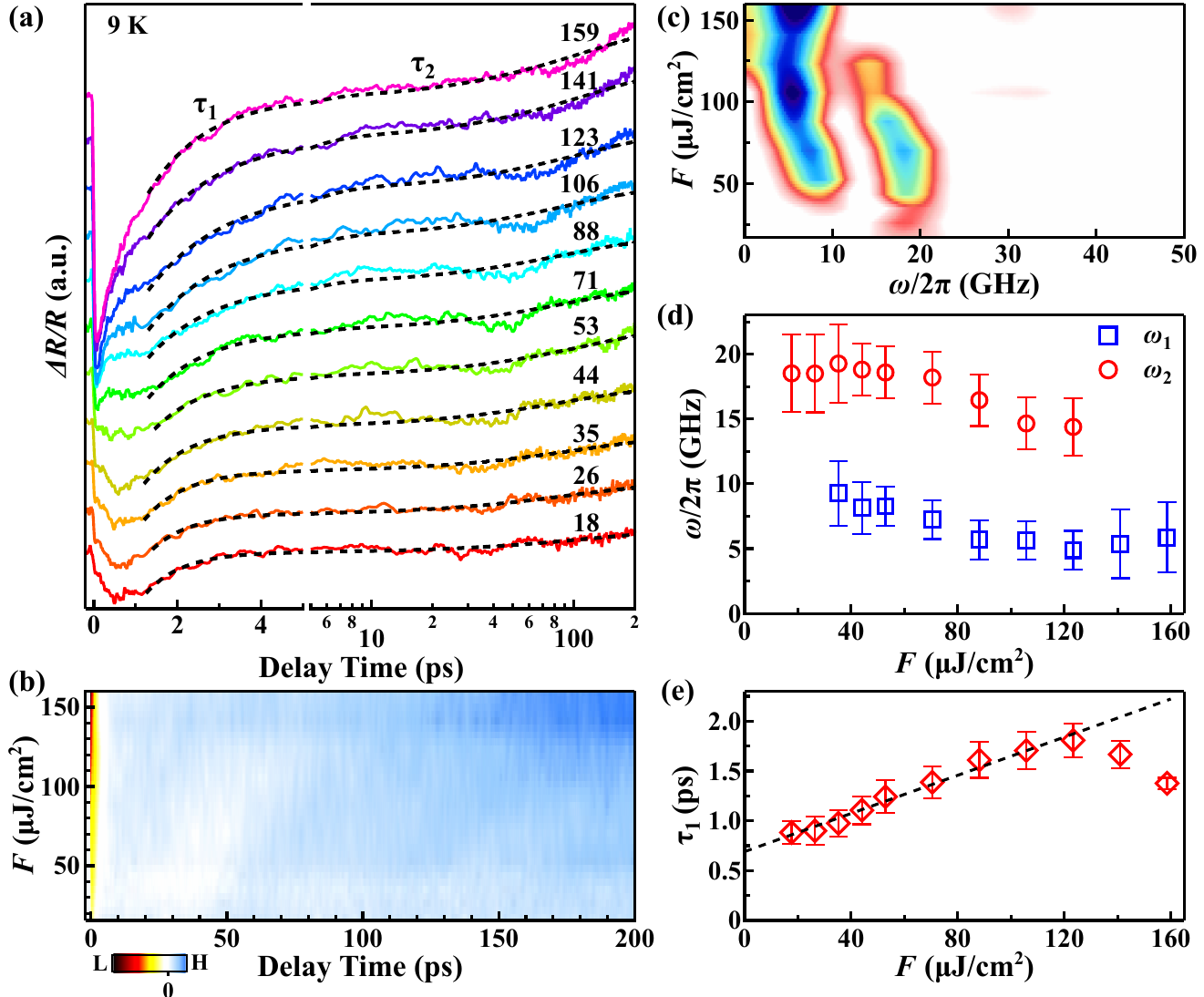}
\caption{(a) Fluence dependence of the $\Delta R/R$ as a function of delay time at 9 K over a long time scales. The black solid lines are Eq. (\blue{1}) fits. (b) 2D pseudocolor map of $\Delta R/R$ as a function of delay time over a fluence range from 18 to 159 {\mjcm}. (c) FFT spectrum color intensity map as a function of frequency and fluence. (d) The derived frequencies of $\omega_1$ and $\omega_2$ as a function of fluence. (e) Fluence dependence of $\tau_1$. The black dashed line represents the linear fit to the low fluence data.}\label{FIG:4}
\end{figure}

In light of the complex magnetic structure revealed by low-temperature magnetization and magnetotransport measurements in {\EIA} \cite{YZhang2020}, indicating the potential for an external field-induced phase transition, we conducted a fluence-dependent measurement at 9 K [Fig. \blue{3(a)}] to explore the characteristics of two modes and investigate potential photon-induced transient phase transitions. The $\Delta R/R$ response exhibits notable variations, with the well-defined reflectivity valley structure around 0.7 ps progressively suppressed and disappearing under high fluence (beyond 123 {\mjcm}). The simplification of the reflectivity curve's lineshape implies the closure of specific relaxation channels as pump fluence surpasses the threshold. Figure \blue{3(b)} presents a 2D pseudocolor map of $\Delta$$R/R$ as a function of pump-probe delay ($x$-axis) and fluence ($y$-axis). The map reveals clear long-period oscillations that exhibit fluence-dependent behavior. For a detailed analysis of oscillations and quasiparticle relaxation, we applied Eq. (\blue{1}) to fit the fluence-dependent transient reflectivity beyond $\sim$1  ps, achieving a good fit, as illustrated by the black dashed lines in Fig. \blue{3(a)}.

FFT was utilized to extract the oscillations, and Fig. \blue{3(c)} presents the FFT spectrum in terms of frequency and fluence. Notably, $\omega_1$ is challenging to discern at low fluence, while $\omega_2$ becomes indiscernible at high fluence. The derived frequencies of $\omega_1$ and $\omega_2$, displayed in Fig. \blue{3(d)}, offer further insights. For fluences below $\sim$71 {\mjcm}, $\omega_2$ remains nearly constant, subtly softening at higher fluence, and becoming undetectable beyond 123 {\mjcm}. In contrast, $\omega_1$, initially indiscernible at low fluence, exhibits gradual softening and even slight abnormal hardening beyond 123 {\mjcm}. In addition, Fig. \blue{3(c)} shows that the FFT amplitude of the $\omega_1$ mode increases with the rise in pump fluence. The derived $\tau_1$ was plotted in Fig. \blue{3(e)} as a function of pump fluence. Initially, $\tau_1$ increases almost monotonically linearly with fluence, but beyond 123 {\mjcm}, it decreases with further fluence, aligning partially with observations in MnBi$_4$Te$_7$, where an increase in relaxation time with fluence was reported \cite{Majchrzak2023}. Calculations suggest that $\omega_1$ could not be a coherent acoustic phonon induced by thermal stress (see details in the Supplemental Material \cite{SupplementalMaterial}). Therefore, we consider that the $\omega_1$ mode might be excited by acoustic waves through a laser-induced magnetoelastic mechanism \cite{TParpiiev2021, YShin2022}. However, further experiments and theoretical models are required to determine the origin of $\omega_1$.

To address the reduced $\tau_1$ and the anomalous behavior of two modes beyond 123 {\mjcm}, we considered two possibilities: firstly, intense pumping causing pronounced heating, elevating the temperature of the illuminated area. Despite the lack of direct low-temperature thermal conductivity measurements \cite{KShinozaki2021}, a preliminary estimation at 9 K was conducted for {\EIA}. The observed lower resistivity in {\EIA} \cite{Goforth2008, Rosa2012, YZhang2020} compared to its newly formed layered compound, $\beta$-{\EIA} \cite{DSWu2023}, along with the higher specific heat in {\EIA} \cite{Rosa2012} than in $\beta$-{\EIA} \cite{DSWu2023}, suggests that the thermal conductivity of {\EIA} possibly exceeds that of $\beta$-{\EIA} \cite{DSWu2023}. Consequently, at 9 K, the estimated thermal conductivity of {\EIA} is approximately 40 W/m$\cdot$K. This estimation aligns with various measurements \cite{Goforth2008, Rosa2012, SRegmi2020, YZhang2020, TSato2020}, including ARPES \cite{SRegmi2020, YZhang2020, TSato2020}, supporting the characterization of {\EIA} as a metal. Considering optical constants \cite{BXu2021} and a worst-case scenario (base temperature of 9 K and excitation density of 159 {\mjcm}), a sample steady-state heat diffusion model \cite{Demsar2006} suggests that the average laser-induced heating in {\EIA} is less than 0.8$^\circ$ and can be neglected. Second, we consider a photon-inducing rapid nonthermal phase transition, such as the quenching of magnetic order \cite{Demsar2006, Stoica2022, YZZhao2023}. High fluence likely disrupts the long-range AFM order, leading to a light-induced transient magnetic phase transition, accompanied by specific relaxation channels. Future work should focus on verifying these transitions using TR-ARPES to better understand the ultrafast changes in the band structure, as well as direct measurements of magnetic order under ultrafast photoexcitation. These studies would provide further insights into the possibility of controlling AFM order and the potential for creating transient magnetic phases in {\EIA}.

\vspace*{-0.3cm}
\section{IV. CONCLUSIONS}
\vspace*{-0.3cm}

In summary, our ultrafast optical spectroscopy study of {\EIA} has provided critical insights into its antiferromagnetic phase transition and the interplay between magnetism and topology. We observed a dramatic change in quasiparticle relaxation dynamics around the N\'{e}el temperature ($T_N$ $\approx$ 16 K), along with the emergence of two distinct low-energy collective modes $\omega_1$ ($\sim$9.9 GHz) and $\omega_2$ ($\sim$21.6 GHz) below $T_N$. Although the origin of the $\omega_1$ mode is currently unknown and cannot be attributed to a coherent acoustic phonon, the $\omega_2$ mode is associated with a magnon, and both exhibit strong temperature dependence, highlighting the intricate coupling between magnetic order and the material's electronic structure. Moreover, high-fluence photoexcitation experiments suggest the possibility of light-induced nonthermal phase transitions, such as the quenching of antiferromagnetic order. This indicates the potential for manipulating magnetic phases in {\EIA} using ultrafast laser pulses, opening new avenues for exploring dynamic magnetic phenomena. Our results establish {\EIA} as a promising platform for studying the interplay between magnetism and topology, with potential applications in quantum information and spintronics. Further investigations, particularly using techniques such as time-resolved and angle-resolved photoemission spectroscopy, will be crucial for precisely characterizing these phase transitions and their implications for topological surface states. These future studies may unlock different opportunities for controlling magnetic order and realizing transient magnetic phases in magnetic topological materials.

\vspace*{-0.3cm}
\section{ACKNOWLEDGMENTS}
\vspace*{-0.3cm}

This work was supported by National Key Research and Development Program of China (Grant No. 2022YFA1604204), the National Natural Science Foundation of China (Grants No. 12074436, U2032204, U22A6005, and 12374163), the Science and Technology Innovation Program of Hunan province (Grant No. 2022RC3068), the Strategic Priority Research Program of the Chinese Academy of Sciences (Grant No. XDB33010000), the Changsha Natural Science Foundation (Grant No. kq2208254), and the Fundamental Research Funds for the Central Universities of Central South University (Grant No. 1053320215412). We are grateful for resources from the High Performance Computing Center of Central South University.


\begin{thebibliography}{99}
\bibitem{MZHasan2010}
M. Z. Hasan and C. L. Kane, Colloquium: Topological insulators, Rev. Mod. Phys. \textbf{82}, 3045 (2010).

\bibitem{SCZhang2011}
X. L. Qi and S. C. Zhang, Topological insulators and superconductors, Rev. Mod. Phys. \textbf{83}, 1057 (2011).

\bibitem{Bernevig2022}
B. A. Bernevig, C. Felser, and H. Beidenkopf, Progress and prospects in magnetic topological materials, Nature \textbf{603}, 41 (2022).

\bibitem{YFXu2020}
Y. F. Xu, L. Elcoro, Z. Song, B. J. Wieder, M. G. Vergniory, N. Regnault, Y. Chen, C. Felser, and B. A. Bernevig, High-throughput calculations of magnetic topological materials, Nature \textbf{586}, 702 (2020).

\bibitem{YFXu2019}
Y. F. Xu, Z. Song, Z. Wang, H. Weng, and X. Dai, Higher-Order Topology of the Axion Insulator $\mathrm{ EuIn_2As_2 }$, Phys. Rev. Lett. \textbf{122}, 256402 (2019).

\bibitem{Goforth2008}
A. M. Goforth, P. Klavins, J. C. Fettinger, and S. M. Kauzlarich, Magnetic Properties and Negative Colossal Magnetoresistance of the Rare Earth Zintl phase $\mathrm{EuIn_2As_2}$, Inorg. Chem. \textbf{47}, 11048 (2008)

\bibitem{Rosa2012}
P. F. S. Rosa, C. Adriano, T. M. Garitezi, R. A. Ribeiro, Z. Fisk, and P. G. Pagliuso, Electron spin resonance of the intermetallic antiferromagnet $\mathrm{EuIn_2As_2}$, Phys. Rev. B \textbf{86}, 094408 (2012).

\bibitem{Riberolles2021}
S. X. M. Riberolles, T. V. Trevisan, B. Kuthanazhi, T. W. Heitmann, F. Ye, D. C. Johnston, S. L. Bud Ko, D. H. Ryan, P. C. Canfield, A. Kreyssig, A. Vishwanath, R. J. McQueeney, L. L. Wang, P. P. Orth, and B. G. Ueland, Magnetic crystalline-symmetry-protected axion electrodynamics and field-tunable unpinned Dirac cones in $\mathrm{EuIn_2As_2}$, Nat. Commun. \textbf{12}, 999 (2021).

\bibitem{JRSoh2023}
J. Soh, A. Bombardi, F. Mila, M. C. Rahn, D. Prabhakaran, S. Francoual, H. M. R{\o}nnow, and A. T. Boothroyd, Understanding unconventional magnetic order in a candidate axion insulator by resonant elastic x-ray scattering, Nat. Commun. \textbf{14}, 3387 (2023).

\bibitem{SRegmi2020}
S. Regmi, M. M. Hosen, B. Ghosh, B. Singh, G. Dhakal, C. Sims, B. Wang, F. Kabir, K. Dimitri, Y. Liu, A. Agarwal, H. Lin, D. Kaczorowski, A. Bansil, and M. Neupane, Temperature-dependent electronic structure in a higher-order topological insulator candidate $\mathrm{EuIn_2As_2}$, Phys. Rev. B \textbf{102}, 165153 (2020).s

\bibitem{TSato2020}
T. Sato, Z. Wang, D. Takane, S. Souma, C. Cui, Y. Li, K. Nakayama, T. Kawakami, Y. Kubota, C. Cacho, T.K. Kim, A. Arab, V.N. Strocov, Y. Yao, and T. Takahashi, Signature of band inversion in the antiferromagnetic phase of axion insulator candidate $\mathrm{EuIn_2As_2}$, Phys. Rev. Res \textbf{2}, 033342 (2020).

\bibitem{YZhang2020}
Y. Zhang, K. Deng, X. Zhang, M. Wang, Y. Wang, C. Liu, J. W. Mei, S. Kumar, E.F. Schwier, K. Shimada, C. Y. Chen, and B. Shen, In-plane antiferromagnetic moments and magnetic polaron in the axion topological insulator candidate $\mathrm{EuIn_2As_2}$, Phys. Rev. B \textbf{101}, 205126 (2020).

\bibitem{MDGong2022}
M. Gong, D. Sar, J. Friedman, D. Kaczorowski, S. Abdel Razek, W. C. Lee, and P. Aynajian, Surface state evolution induced by magnetic order in axion insulator candidate $\mathrm{EuIn_2As_2}$, Phys. Rev. B \textbf{106}, 125156 (2022).

\bibitem{DNBasov2011}
D. N. Basov, R. D. Averitt, D. van der Marel, M. Dressel, and K. Haule, Electrodynamics of correlated electron materials, Rev. Mod. Phys. \textbf{83}, 471 (2011).

\bibitem{SXZhu2021}
S. X. Zhu, C. Zhang, Q. Y. Wu, X. F. Tang, H. Liu, Z. T. Liu, Y. Luo, J. J. Song, F. Y. Wu, Y. Z. Zhao, S. Y. Liu, T. Le, X. Lu, H. Ma, K. H. Liu, Y. H. Yuan, H. Huang, J. He, H. Y. Liu, Y. X. Duan, and J. Q. Meng, Temperature evolution of quasiparticle dispersion and dynamics in semimetallic 1$\textit{T}$-$\mathrm{TiTe_2}$ via high-resolution angle-resolved photoemission spectroscopy and ultrafast optical pump-probe spectroscopy, Phys. Rev. B \textbf{103}, 115108 (2021).

\bibitem{JBQi2010}
J. Qi, X. Chen, W. Yu, P. Cadden-Zimansky, D. Smirnov, N. H. Tolk, I. Miotkowski, H. Cao, Y. P. Chen, Y. Wu, S. Qiao, and Z. Jiang, Ultrafast carrier and phonon dynamics in $\mathrm{Bi_2Se_3}$ crystals, Appl. Phys. Lett. \textbf{97}, 182102 (2010).

\bibitem{CZhang2022}
C. Zhang, Q. Y. Wu, W. S. Hong, H. Liu, S. X. Zhu, J. J. Song, Y. Z. Zhao, F. Y. Wu, Z. T. Liu, S. Y. Liu, Y. H. Yuan, H. Huang, J. He, S. L. Li, H. Y. Liu, Y. X. Duan, H. Q. Luo, and J. Q. Meng, Ultrafast optical spectroscopy evidence of pseudogap and electron-phonon coupling in an iron-based superconductor $\mathrm{ KCa_2Fe_4As_4F_2 }$, Sci. China-Phys. Mech. Astron. \textbf{65}, 237411 (2022).

\bibitem{QYWu2023A}
Q. Y. Wu, C. Zhang, Z. Z. Li, W. S. Hong, H. Liu, J. J. Song, Y. Z. Zhao, Y. H. Yuan, B. Chen, X. Q. Ye, S. L. Li, J. He, H. Y. Liu, Y. X. Duan, H. Q. Luo, and J. Q. Meng, Hidden nematic fluctuation in the triclinic $\mathrm{(Ca_{0.85}La_{0.15})_{10}(Pt_3As_8)(Fe_2As_2)_5}$ superconductor revealed by ultrafast optical spectroscopy, Phys. Rev. B \textbf{108}, 205136 (2023).

\bibitem{QYWu2023B}
Q. Y. Wu and J. Q. Meng, Ultrafast optical spectroscopy of FeAs-based superconductors, Sci. Sin.: Phys. Mech. Astron. \textbf{53}, 127408 (2023)

\bibitem{Kabanov1999}
V. V. Kabanov, J. Demsar, B. Podobnik, and D. Mihailovic, Quasiparticle relaxation dynamics in superconductors with different gap structures: Theory and experiments on $\mathrm{YBa_2Cu_3O_{7-\delta}}$, Phys. Rev. B \textbf{59}, 1497 (1999).

\bibitem{Demsar2006}
J. Demsar, J. L. Sarrao, and A. J. Taylor, Dynamics of photoexcited quasiparticles in heavy electron compounds, J. Phys.: Condens. Matter \textbf{18}, R281 (2006).

\bibitem{YZZhao2023}
Y. Z. Zhao, Q. Y. Wu, C. Zhang, B. Chen, W. Xia, J. J. Song, Y. H. Yuan, H. Liu, F. Y. Wu, X. Q. Ye, H. Y. Zhang, H. Huang, H. Y. Liu, Y. X. Duan, Y. F. Guo, J. He, and J. Q. Meng, Coupling of optical phonons with Kondo effect and magnetic order in the antiferromagnetic Kondo-lattice compound $\mathrm{CeAuSb_2}$, Phys. Rev. B \textbf{108}, 075115 (2023).

\bibitem{KSBurch2008}
K. S. Burch, E. E. M. Chia, D. Talbayev, B. C. Sales, D. Mandrus, A. J. Taylor, and R. D. Averitt, Coupling between an Optical Phonon and the Kondo Effect, Phys. Rev. Lett. \textbf{100}, 026409 (2008).

\bibitem{YPLiu2020}
Y. P. Liu, Y. J. Zhang, J. J. Dong, H. Lee, Z. X. Wei, W. L. Zhang, C. Y. Chen, H. Q. Yuan, Y. F. Yang, and J. Qi, Hybridization Dynamics in $\mathrm{CeCoIn_5}$ Revealed by Ultrafast Optical Spectroscopy, Phys. Rev. Lett. \textbf{124}, 057404 (2020).

\bibitem{JQi2013}
J. Qi, T. Durakiewicz, S. A. Trugman, J.X. Zhu, P. S. Riseborough, R. Baumbach, E. D. Bauer, K. Gofryk, J. Q. Meng, J. J. Joyce, A. J. Taylor, and R. P. Prasankumar, Measurement of Two Low-Temperature Energy Gaps in the Electronic Structure of Antiferromagnetic $\mathrm{USb_2}$ Using Ultrafast Optical Spectroscopy, Phys. Rev. Lett. \textbf{111}, 057402 (2013).

\bibitem{Rothwarf1967}
A. Rothwarf and B.N. Taylor, Measurement of Recombination Lifetimes in Superconductors, Phys. Rev. Lett. \textbf{19}, 27 (1967).

\bibitem{VIyer2018}
V. Iyer, Y. P. Chen, and X. F. Xu, Ultrafast Surface State Spin-Carrier Dynamics in the Topological Insulator $\mathrm{Bi_2Te_2Se}$, Phys. Rev. Lett. \textbf{121}, 026807 (2018).

\bibitem{PSharma2022}
P. Sharma, A. Bhardwaj, R. Sharma, V. P. S. Awana, T. N. Narayanan, K. V. Raman, and M. Kumar, Comprehensive Study of the topological Surface States through Ultrafast Pump-probe Spectroscopy, J. Phys. Chem. C \textbf{126}, 11138(2022).

\bibitem{YLi2021}
Y. Li, H. Deng, C. Wang, S. Li, L. Liu, C. Zhu, K. Jia, Y. Sun, X. Du, X. Yu, T. Guan, R. Wu, S. Zhang, Y. Shi, and H. Mao, Surface and electronic structure of antiferromagnetic axion insulator candidate $\mathrm{EuIn_2As_2}$, Acta Phys. Sin. \textbf{70}, 186801 (2021).

\bibitem{SupplementalMaterial}
See Supplemental Material at \blue{****} for additional experimental results and a supporting data analysis, which includes Refs.\cite{YZhang2020, DLim2003, MTakahara2012, DWang2007, FHYu2020, BXu2021, QiongWu2023}.

\bibitem{DLim2003}
D. Lim, R. D. Averitt, J. Demsar, A. J. Taylor, N. Hur, and S. W. Cheong, Coherent acoustic phonons in hexagonal manganite $\mathrm{LuMnO_3}$, Appl. Phys. Lett. \textbf{83}, 4800 (2003).

\bibitem{MTakahara2012}
M. Takahara, H. Jinn, S. Wakabayashi, T. Moriyasu, and T. Kohmoto, Observation of coherent acoustic phonons and magnons in an antiferromagnet NiO, Phys. Rev. B \textbf{86}, 094301 (2012).

\bibitem{DWang2007}
D. Wang, A. Cross, G. Guarino, S. Wu, R. Sobolewski, and A. Mycielski, Time-resolved dynamics of coherent acoustic phonons in CdMnTe diluted-magnetic single crystals, Appl. Phys. Lett. \textbf{90}, 211905 (2007).

\bibitem{FHYu2020}
F. H. Yu, H. M. Mu, W. Z. Zhuo, Z. Y. Wang, Z. F. Wang, J. J. Ying, and X. H. Chen, Elevating the magnetic exchange coupling in the compressed antiferromagnetic axion insulator candidate $\mathrm{EuIn_2As_2}$, Phys. Rev. B \textbf{102}, 180404(R) (2020).

\bibitem{BXu2021}
B. Xu, P. Marsik, S. Sarkar, F. Lyzwa, Y. Zhang, B. Shen, and C. Bernhard, Infrared study of the interplay of charge, spin, and lattice excitations in the magnetic topological insulator $\mathrm{EuIn_2As_2}$, Phys. Rev. B \textbf{103}, 245101 (2021).

\bibitem{QiongWu2023}
Q. Wu, T. Hu, D. Wu, R. Li, L. Yue, S. Zhang, S. Xu, Q. Liu, T. Dong, and N. Wang, Spin dynamics in the axion insulator candidate $\mathrm{EuIn_2As_2}$, Phys. Rev. B \textbf{107}, 174411 (2023).

\bibitem{HLi2019}
H. Li, S. Y. Gao, S. F. Duan, Y. F. Xu, K. J. Zhu, S. J. Tian, J. C. Gao, W. H. Fan, Z. C. Rao, J. R. Huang, J. J. Li, D. Y. Yan, Z. T. Liu, W. L. Liu, Y. B. Huang, Y. L. Li, Y. Liu, G. B. Zhang, P. Zhang, T. Kondo, S. Shin, H. C. Lei, Y. G. Shi, W. T. Zhang, H. M. Weng, T. Qian, and H. Ding, Dirac Surface States in Intrinsic Magnetic Topological Insulators $\mathrm{EuSn_2As_2}$ and $\mathrm{MnBi_{2n}Te_{3n+1}}$, Phys. Rev. X \textbf{9}, 041039 (2019).

\bibitem{Majchrzak2023}
P. E. Majchrzak, Y. Liu, K. Volckaert, D. Biswas, C. Sahoo, D. Puntel, W. Bronsch, M. Tuniz, F. Cilento, X. Pan, Q. Liu, Y. P. Chen, and S. Ulstrup, van der Waals engineering of ultrafast carrier dynamics in magnetic heterostructures, Nano Lett. \textbf{23}, 414 (2023).

\bibitem{YSWang2023}
Y. S. Wang, B. Chen, Z. T. Liu, X. Guo, S. You, Z. Wang, H. Xie, T. Niu, J. Q. Meng, and H. Huang, In-plane electron-phonon coupling anisotropy and multiple charge density wave orders in the superconductor $\mathrm{Bi_2Rh_3Se_2}$, Phys. Rev. B \textbf{108}, 045118 (2023).

\bibitem{RVYusupov2008}
R. V. Yusupov, T. Mertelj, J. H. Chu, I. R. Fisher, and D. Mihailovic, Single-Particle and Collective Mode Couplings Associated with 1- and 2-Directional Electronic Ordering in Metallic \textit{R}$\mathrm{Te_3}$ (\textit{R} = Ho , Dy , Tb), Phys. Rev. Lett. \textbf{101}, 246402 (2008).

\bibitem{MFCollins1989}
M. F. Collins, $Magnetic Critical Scattering$ (Oxford University Press, Oxford, U.K., 1989).

\bibitem{Afanasiev2021}
D. Afanasiev, J. R. Hortensius, M. Matthiesen, S. Ma\~{n}as-Valero, M. \v{S}i\v{s}kins, M. Lee, E. Lesne, H. S. J. van der Zant, P. G. Steeneken, B. A. Ivanov, E. Coronado, and A. D. Caviglia, Controlling the anisotropy of a van der Waals antiferromagnet with light, Sci. Adv. \textbf{7}, eabf3096 (2021).

\bibitem{Kalashnikova2007}
A. M. Kalashnikova, A. V. Kimel, R. V. Pisarev, V. N. Gridnev, A. Kirilyuk, and T. Rasing, Impulsive Generation of Coherent Magnons by Linearly Polarized Light in the Easy-Plane Antiferromagnet $\mathrm{FeBO_3}$, Phys. Rev. Lett. \textbf{99}, 167205 (2007).

\bibitem{Nishitani2013}
J. Nishitani, T. Nagashima, and M. Hangyo, Terahertz radiation from antiferromagnetic MnO excited by optical laser pulses, Appl. Phys. Lett. \textbf{103}, 081907 (2013).

\bibitem{TParpiiev2021}
T. Parpiiev, A. Hillion, V. Vlasov, V. Gusev, K. Dumesnil, T. Hauet, S. Andrieu, A. Anane, and T. Pezeril, Ultrafast strain excitation in highly magnetostrictive terfenol: Experiments and theory, Phys. Rev. B \textbf{104}, 224426 (2021).

\bibitem{YShin2022}
Y. Shin, M. Vomir, D. Kim, P. C. Van, J. Jeong, and J. Kim, Quasi-static strain governing ultrafast spin dynamics, Communications Physics \textbf{5}, 56 (2022).

\bibitem{KShinozaki2021}
K. Shinozaki, Y. Goto, K. Hoshi, R. Kiyama, N. Nakamura, A. Miura, C. Moriyoshi, Y. Kuroiwa, H. Usui, and Y. Mizuguchi, Thermoelectric Properties of the As/P-Based Zintl Compounds $\mathrm{ EuIn_2As_{2-x}P_x}$ (x = 0-2) and $\mathrm{SrSn_2As_2}$, ACS Appl. Energy Mater. \textbf{4}, 5155 (2021).

\bibitem{DSWu2023}
D. S. Wu, S. H. Na, Y. J. Li, X. B. Zhou, W. Wu, Y. T. Song, P. Zheng, Z. Li, and J. L. Luo, Single-crystal growth, structure and thermal transport properties of the metallic antiferromagnet Zintl-phase $\mathrm{ \beta}$-$\mathrm{EuIn_2As_2 }$, Phys. Chem. Chem. Phys. \textbf{26}, 8695 (2024).

\bibitem{Stoica2022}
V. A. Stoica, D. Puggioni, J. Zhang, R. Singla, G. L. Dakovski, G. Coslovich, M. H. Seaberg, M. Kareev, S. Middey, P. Kissin, R. D. Averitt, J. Chakhalian, H. Wen, J. M. Rondinelli, and J. W. Freeland, Magnetic order driven ultrafast phase transition in $\mathrm{NdNiO_3}$, Phys. Rev. B \textbf{106}, 165104 (2022).

\end{thebibliography}
\end{document}